\documentclass[footinbib,twocolumn,showpacs,pra,preprintnumbers,amsmath,amssymb,aps,floatfix]{revtex4}
\usepackage{graphicx}
\usepackage{dcolumn}
\usepackage{bm}
\begin{document}
\title{ Entanglement entropy and the determination of an unknown quantum state}
\newcommand{\e}{{\rm e}}
\newcommand{\veps}{\varepsilon}
\newcommand{\lgl}{\langle}
 \newcommand{\rgl}{\rangle}
\renewcommand{\d}{{\rm d}}
\newcommand{\Vh}[1]{\hat{#1}}
\newcommand{\Aa}{A^1_{\epsilon}}
\newcommand{\Ab}{A^{\epsilon}_L}
\newcommand{\Ae}{A_{\epsilon}}
\newcommand{\finn}[1]{\phi^{\pm}_{#1}}
\newcommand{\ea}{e^{-|\alpha|^2}}
\newcommand{\eb}{\frac{e^{-|\alpha|^2} |\alpha|^{2 n}}{n!}}
\newcommand{\ebbb}{\frac{e^{-3|\alpha|^2} |\alpha|^{2 (l+n+m)}}{l!m!n!}}
\newcommand{\ass}{\alpha}
\newcommand{\as}{\alpha^*}
\newcommand{\fb}{\bar{f}}
\newcommand{\gb}{\bar{g}}
\newcommand{\la}{\lambda}
\newcommand{\sz}{\hat{s}_{z}}
\newcommand{\sy}{\hat{s}_y}
\newcommand{\sx}{\hat{s}_x}
\newcommand{\sio}{\hat{\sigma}_0}
\newcommand{\six}{\hat{\sigma}_x}
\newcommand{\siz}{\hat{\sigma}_{z}}
\newcommand{\siy}{\hat{\sigma}_y}
\newcommand{\vhsig}{\vec{hat{\sigma}}}
\newcommand{\hsig}{\hat{\sigma}}
\newcommand{\hI}{\hat{I}}
\newcommand{\hone}{\hat{1}}
\newcommand{\hs}{\hat{s}}
\newcommand{\ha}{\hat{a}}
\newcommand{\hH}{\hat{H}}
\newcommand{\hU}{\hat{U}}
\newcommand{\hA}{\hat{A}}
\newcommand{\hB}{\hat{B}}
\newcommand{\hC}{\hat{C}}
\newcommand{\hD}{\hat{D}}
\newcommand{\hV}{\hat{V}}
\newcommand{\hW}{\hat{W}}
\newcommand{\hK}{\hat{K}}
\newcommand{\hX}{\hat{X}}
\newcommand{\hM}{\hat{M}}
\newcommand{\hN}{\hat{N}}
\newcommand{\hO}{\hat{O}}
\newcommand{\te}{\theta}
\newcommand{\vze}{\vec{\zeta}}
\newcommand{\vet}{\vec{\eta}}
\newcommand{\vx}{\vec{\xi}}
\newcommand{\vc}{\vec{\chi}}
\newcommand{\hro}{\hat{\rho}}
\newcommand{\vro}{\vec{\rho}}
\newcommand{\hR}{\hat{R}}
\newcommand{\half}{\frac{1}{2}}
\renewcommand{\d}{{\rm d}}
\renewcommand{\top }{ t^{\prime } }
\newcommand{\oz}{{(0)}}
\newcommand{\sint}{{\rm si}}
\newcommand{\cint}{{\rm ci}}
\newcommand{\de}{\delta}
\newcommand{\ep}{\varepsilon}
\newcommand{\De}{\Delta}
\newcommand{\eps}{\varepsilon}
\newcommand{\si}{\hat{\sigma}}
\newcommand{\om}{\omega}
\newcommand{\tr}{{\rm tr}}
\newcommand{\gam}{\gamma ^{(0)}}
\newcommand{\pe}{\prime}
\newcommand{\BEQ}{\begin{equation}}
\newcommand{\EEQ}{\end{equation}}
\newcommand{\BEA}{\begin{eqnarray}}
\newcommand{\EEA}{\end{eqnarray}}
\newcommand{\sph}{spin-$\frac{1}{2}$ }
\newcommand{\ad}{\hat{a}^{\dagger}}
\newcommand{\add}{\hat{a}}
\newcommand{\spp}{\hat{\sigma}_+}
\newcommand{\smm}{\hat{\sigma}_-}
\newcommand{\fin}[1]{|\phi^{\pm}_{#1}\rangle}
\newcommand{\finp}[1]{|\phi^{+}_{#1}\rangle}
\newcommand{\finm}[1]{|\phi^{-}_{#1}\rangle}
\newcommand{\lfin}[1]{\langle \phi^{\pm}_{#1}|}
\newcommand{\lfinp}[1]{\langle \phi^{+}_{#1}|}
\newcommand{\lfinm}[1]{\langle \phi^{-}_{#1}|}
\newcommand{\lfinn}[1]{\langle\phi^{\pm}_{#1}|}
\newcommand{\z}{\cal{Z}}
\newcommand{\RI}{\hat{{\cal{R}}}_{0}}
\newcommand{\Rt}{\hat{{\cal{R}}}_{\tau}}
\newcommand{\nn}{\nonumber}

\author{Gerardo Aquino$^1$}
\author{Filippo Giraldi$^2$}\email{filgi@libero.it} 
\affiliation{$^1$Max-Planck-Institut f\"ur Physik komplexer Systeme-
N\"othnitzer Str. 38
01187 Dresden, Germany}

\affiliation{$^2$Gruppo Nazionale per la  Fisica Matematica, GNFM-INdAM, Via
Madonna del Piano 10, I-50019 Sesto Fiorentino (FI), Italy}

\begin{abstract}
An initial unknown  quantum state can be determined
with a single measurement apparatus by letting it  interact
 with an auxiliary, ``Ancilla",  system  as proposed by Allahverdyan, Balian
and Nieuwenhuizen [Phys. Rev. Lett. 92, 120402 (2004)]. In the case of  two qubits,
this procedure allows to reconstruct the initial state of the qubit of interest $S$ by measuring
three commuting observables and therefore by means of a single apparatus, for the total system $S+A$ at a later time.
 The determinant of the
matrix of the linear transformation connecting the measurements of three commuting observables
at time $t>0$ to the components of the polarization vector of $S$ at time $t=0$ is used 
as an indicator of the reconstructability of the initial state of the system
$S$. We show that a connection between the entanglement entropy of the total
system $S+A$ and such a determinant exists,
and that for a pure state a vanishing entanglement 
individuates, without a need for any  measurement,
 those intervals of time for which the reconstruction procedure  is least
efficient. This  property remains valid for a generic dimension of $S$.
In the case of a mixed state
this connection is lost.

\end{abstract}
\maketitle

{\it Introduction.}
The determination of the unknown state of a quantum system is one
of the most important issues in the field of quantum
information \cite{chuang,hillery,bennett}.
For a qubit the estimation of the density matrix involves the measurement of
three non-commuting observables, i.e.  three successive Stern-Gerlach
measurements performed along  three orthogonal directions are necessary to
determine the components of the Bloch polarization vector $\vec{\rho}$ which determines the state of  $S$.
 In each measurement in fact,  the other two components are destroyed.
Recently \cite{teo}, based on a modification of an idea originally introduced
in  \cite{ariano}, a procedure was proposed to bypass this limitation
by  coupling the system to an ancilla system $A$ whose initial state is known.
Starting from a factorized condition, a measurement of three commuting
observables at time $t$  in the space of the
compound system $S+A$
allows  to reconstruct the state of the system of interest $S$ at time
zero. 
This is feasible if for the respective Hilbert spaces:  $dim\mathcal{H}_ A \geq dim \mathcal{H}_ S$ and  if the  interaction
intertwines  the two systems
so as to give non-zero determinant for the matrix connecting the measured
values of the three observables at time $t$ to the components of  the vector $\vec{\rho}$
  that individuates the state of  $S$ at time $t=0$.
This procedure requires just on instance of measurement, i.e. one single
apparatus (e.g.  simultaneously measuring the $z$-components of the
Spins of $S$ and $A$ and their product, in the case of $S$ and $A$ being two qubits)
and is therefore more economical and  was recently implemented experimentally
in \cite{du}.\\
The procedure extends
 to a generic dimension
of $S$,  as explained in \cite{teo}, by  considering two commuting observables,
 one pertaining to $S$ and the other to $A$
  and evaluating, in repeated experiments, the
probabilities $P_{i j}$
to have as outcomes  the $i^{th}$ eigenvalue of the first observable
and the $j^{th}$ for the second one.
A linear mapping between such probabilities and the initial density matrix of
$S$, ensues.
This is  expected to be invertible provided that
the  number of distinct eigenvalues of both 
the observables is (at least) equal to the dimension of $\mathcal{H}_S$,
this implies the above mentioned costraint on the dimension of
$\mathcal{H}_A$. The particular case of a spin-1/2 particle
coupled to a laser cavity field, described by the Jaynes-Cummings hamiltonian,
important for possible experimental implementations,
was considered in \cite{mine1,bahar2}.

In this article  we answer the question of how, fixed a coupling between $S$
and $A$, the entanglement measure  provides
information on the feasibility and efficiency of the procedure.
We analyze 
the case of $S$ and $A$ being
two qubits, and then  generalize
the arguments to a generic dimension of $S$ and $A$.

 
 {\it Two by two density matrix.}
Let us consider   a \sph  $S$ interacting through a
generic  time independent Hamiltonian $\hat{H}$ with a second 
\sph: the ancilla system $ A$. The total system $S+A$ is
 set in the following  initial  state:
\BEQ \label{initial}
\hro_T(0)=\hro_S (0)\otimes \hro_A(0)=\frac{\hone+\vec{\rho}\cdot \vec{\hat{\sigma}}}{2}\otimes
\frac{\hone+\lambda \hat{s}_3 }{2}
\EEQ
 where the components 
$\hat{\sigma}_i$, with $i=1,2,3$, are the Pauli one half spin operators
acting on the Hilbert space $\mathcal{H}_S$ of the spin of interest, and
$\hat{s}_j$  are the analogous   operators acting on
the Hilbert spin space  $\mathcal{H}_A$ of the Ancilla. 
 In  the case
of an initial pure state, $\hro_T^2 (0)=\hro_T(0)$,which means
$|\vec{\rho}|^2=1$ and $\lambda=\pm1$.
Since the Hamiltonian $\hat{H}$ is time independent  the time evolution operator  $\hat{U}(t)=e^{-i \hat{H}
t}$ is unitary, this implies that if initially the system is described by a
pure quantum state, the quantum state remains pure at any following time.

 Furthermore, using the properties of the evolution operator, the expectation
value of a general operator $O$ at time $t$, acting on the
Hilbert  space $\mathcal{H}_S\otimes
\mathcal{H}_A$,  can be easily calculated as:
 \BEQ\label{expctvalO}
 \langle \hat{O}\rangle_t= Tr_A [ e^{-i \hat{H}t}\hro_T(0) e^{i \hat{H}t}\hat{O }], \end{equation}
from which it descends that: $\big(\mathfrak{i}\big)$  $\langle O (t) \rangle$ is just a \emph{linear} function of
$\rho_1, \rho_2$ and $\rho_3$, the parameters describing the initial
quantum state of the spin of interest and therefore $\big(\mathfrak{ii}\big)$ \emph{no quadratic} term,
$a_{ij} \rho_i
\rho_j$, appears in the expectation value of a generic operator $\hat{O}$.

 We consider, now, two general observables whose operators are: $\hat{O}_S$, related to the spin of
interest, $\hat{O}_A$, related to the Ancilla spin, and the  observable
related to the operator $\hat{O}_S \otimes \hat{O}_A$. We are interested in the
determinant of the $3 \times 3 $ matrix $M$, defined by the
relation:
\BEQ
\label{basic}
 \vec{p}(t)=\Omega(t)\cdot \vec{\rho}+ \vec{k}(t),
\EEQ
 where $\vec{y}$ represents the column
matrix of elements: $\langle \hat{O}_S (t) \rangle$, $\langle \hat{O}_A
(t) \rangle$ and $\langle \hat{O}_S (t) \otimes \hat{O}_A(t) \rangle$, $\vec{\rho}$ is the polarization vector in Eq. (\ref{initial}) 
 and $\vec{k}$ is a time
dependent column matrix. If the state is initially factorized, according to Eq. (\ref{initial}) the determinant of the
matrix $M$ at time $t=0$ is obviously zero. If at a generic later time $t$ the determinant does not vanish,
Eq. (\ref{basic}) can be inverted and therefore the
initial state of the spin of interest can be derived from the
expectation values measured at time $t$. 

%
%

Since only in the case of pure total state a unique definition of entanglement
measure exists, we consider separately the cases of initial pure total state
and initial  mixed total state, even though in  the more general protocol
the  initial state of the system of interest is totally unknown,  including
therefore its pure or mixed nature.

We wish to show that in the case of initial pure state 
a correspondence exists between the vanishing quantum Entanglement and
the vanishing determinant of $M$.
We will see that
  a vanishing entanglement gives information
about the intervals of time where the reconstruction of the initial state
is least efficiently  implemented.

{\it{The case of  initial pure state.}} 
Let us consider the total system  in a pure
 state at time $t=0$, i.e. the initial total density matrix is given by the
expression of Eq. (\ref{initial}) with
$|\vec{\rho}|^2=1$  and $\lambda=\pm1$. Obviously, the system
will evolve in time through pure states: $\hro_T^2 (t)=\hro_T(t)$.
 Let us assume that the quantum
entanglement of the \emph{pure state} $\hro_T(t)$, \emph{vanishes} at a
certain instant $t^{\ast}$, this way the
system is described by the quantum state ket
$|\alpha \rangle_{t^*}
|\beta \rangle_{t^*}$, thus, the expectation value of the operator
$ \hat{O}_S \otimes \hat{O}_A $ is the product of the expectation values of operators $\hat{O}_S$ and $\hat{O}_A$:
\begin{equation}
\langle \hat{O}_S \otimes \hat{O}_A  \rangle_{t^*}=
\langle \hat{O}_S   \rangle_{t^*} \langle \hat{O}_S
  \rangle_{t^*}. \label{O1O2expctval}
\end{equation}
which obviously means that the mapping (\ref{basic}) is not invertible
since the state of system $S$ is described by three independet parameters
while the independent components of vector $\vec{p}$ are only two.
More in detail, following Eq. (\ref{expctvalO}), we know that
the expectation values of the operators $\hat{O}_S$ and $\hat{O}_A$ are
described by the following linear relations:
\begin{equation}
\label{linearRelO}
\langle \hat{O}_{S(A)}  \rangle_{t^*}=\vec{\gamma}_{S(A)}\cdot
\vec{\rho} + \delta_{S(A)}, 
\end{equation}
where $\gamma^i_S\equiv M_{1,i} ,\gamma^{i}_{A}\equiv
M_{2,i} $ and $ \delta_{S(A)}\equiv k_{1(2)} $. According to Eq.
(\ref{O1O2expctval}), we obtain the following expression for the
expectation value of $\hat{O}_S \otimes \hat{O}_A$:
\begin{equation}
\label{O1O2expvall}
\langle \hat{O}_S\otimes \hat{O}_A \rangle_{t^*}=\gamma^{i}_S \gamma^{j}_A
\rho_i \rho_j + \delta_{S} \gamma^{i}_A \rho_i+\delta_{A} 
\gamma^{i}_S \rho_i+\delta_{S} \delta_{A}. 
\end{equation}
where sum of repeated indexes is implied.
According to the observation $\big(\mathfrak{i}\big)$, we have
$\gamma^{i}_S \gamma^{j}_A=0$ for every $i,j=1,2,3$; thus, either
$\gamma^{i}_S=0$,  or $\gamma^{i}_A=0$, or both $\gamma^{i}_S=\gamma^{i}_A=0$, for every $i=1,2,3$.
In any case, at least one of the first two rows of the matrix
$\Omega $ vanishes and the third row is proportional to the non vanishing row; for example, in case the second row
vanishes, the third row is $\delta_{S}$ times the first row. This way,
we have confirmed that the determinant vanishes:
$\Delta(t^*)\equiv det[\Omega(t^{\ast})]=0$.

Now, we study the time derivative of the determinant through the
property: 
\BEQ
\label{detderiv}
\frac{d}{dt} \Delta(t) \equiv\frac{d}{dt} det[\Omega(t)]=  \Omega_{i,j}\frac{d}{dt}{[\Omega(t)]_{i,j}},
\EEQ
 where
$\left[\Omega(t)\right]_{i,j}$ denotes  matrix element of row $i$ and
column $j$, and $\Omega_{i,j}$ is the corresponding cofactor. Since one
of the first two rows of $\Omega $ vanishes, and
since the third row is proportional to the non vanishing row, every
cofactor of the matrix $\Omega $ vanishes, which
means that, \emph{when} the quantum Entanglement vanishes, the time
derivative of the determinant vanishes too, i.e.  $[
d\Delta(t)/dt]_{t=t^{\ast}}=0$.


 For a generic dimension
$N$ of $S$ and $M$ of $A$  as explained in \cite{teo}  one considers two
commuting observables with nondegenerate spectrum,one  pertaining to $S$ and the other to $A$ which read
in their spectral decomposition as
$\hat{O}_S=\sum_{i=1}^{N} s_i \hat{s}_i$  and $\hat{O}_A=\sum_{j=1}^{M} a_j\hat{a}_j$ 
  One then evaluates, in repeated experiments, the
 probabilities 
\BEQ
P_{i j}=\langle \hs_i \otimes \ha_j \rangle
\EEQ
to have as outcomes  the $i^{th}$ eigenvalue of the first observable
and the $j^{th}$ for the second one.
A linear mapping between the $N\times M$ component  vector ${\bf p}$ such that $p_{\alpha}=P_{ij}$,
with $\alpha=\{ij\}$ ,  and the initial density matrix of $S$, ensues
As already mentioned, this mapping is  expected to be invertible provided that
the  number of distinct eigenvalues of both 
the observables is (at least) equal to the dimension of $\mathcal{H}_S$,
which implies the constraint $M\geq N$.
We consider here the case where the Ancilla system has the same dimension
of $S$,i.e. $M=N$.
In this case the mapping reads:
\BEQ
\label{restrict}
{\bf p}(t)=\Omega(t)\cdot {\boldsymbol{\rho}}+ {\bf{k}}(t)
\EEQ
where ${\boldsymbol{\rho}}$ is the $N^2-1$ components vector, containing all the
independent parameters characterizing the state of system $S$ at time $t=0$
and $\Omega$ is a $(N^2-1)\times( N^2-1)$ square matrix.  Vector ${\bf p}$
in Eq.(\ref{restrict}) is therefore restricted  to ist first $N^2-1$
components, i.e. the last component $p_{N^2}=P_{NN}=\langle \hs_N \otimes \ha_N \rangle$, being fixed by
normalization, is omitted.
 
For convenience we redefine vector ${\bf p}$ components in Eq. (\ref{restrict})to be 
 \BEQ
p_{\alpha}(t)=\left\{
\rule{0 cm}{0.65 cm}
\begin{array}{cc}
\langle \hs_{i} \otimes \hat{1}\rangle_t \;\; &i=\alpha=1,\dots,N\\
\langle \hat{1}\otimes \ha_{j}\rangle_t  \;\;&j=2,\dots,N,\;\;\alpha=N+j-1\\
\langle \hs_i\otimes \ha_j\rangle_t  \;\;&i,j \geq 2, \;\; \alpha=2N, \dots,N^2-1\\
\end{array}
\right.
\EEQ
which amounts to a linear combination rearrangement of the original
$N^2-1$  components of ${\bf p}$ leaving therefore the determinant
of
$\Omega$ unchanged.
Observation ({\it i}) is still valid for this case,
so  a generic component $p_{\alpha}$ with $\alpha<N$ can be written as:
\BEQ
p_{\alpha}=\langle\hs_{\alpha} \otimes \hat{1}_A \rangle_t= \lambda^{\alpha}_j \rho_j
+\delta^{\alpha}_S
\EEQ
The same is true for the components with $N+1 \leq \alpha \leq 2N$
\BEQ
p_{\alpha}=\langle\hat{1}_S \otimes \ha_{\alpha} \rangle_t= \lambda^{\alpha}_{j} \rho_{j}
+\delta^{\alpha}_S
\EEQ
Again if the initial state is pure and if the quantum entanglement vanishes at
time $t^*$, then, at this time, the 
system is described by the quantum state ket
$|\alpha \rangle
|\beta \rangle$ and therefore
\BEA
\langle \hs_n \otimes \ha_m \rangle_{t^{*}}&=&\langle \hs_n
\rangle_{t^*}\langle\ha_m\rangle_{t^*}=\\
\nonumber &=& \lambda^n_{j} \lambda^m_{k}\rho_j \rho_k+
\delta^m_A \lambda^n_j\rho_j +\delta^n_S \lambda^m_k \rho_k +\delta^m_A \delta^n_S.
\EEA
One easily realizes that the argument used for the two qubits case
applies again in similar fashion.
In fact following observation ({\it i}) $\lambda^n_j \lambda^m_k=0$ for all $j,k$ ,
which means that either $\lambda^n_j=0$ for every $j$ or $\lambda^m_k=0$ 
for every $ k$ or $\lambda^n_j=\lambda^m_k=0$  for all $j,k$.
But this implies that either  row $n$ or row $m$ of matrix $\Omega$ vanishes and that
the row corrsesponding to $\langle \hs_n \otimes \ha_m \rangle_{t^*}$ is
proportional to the non vanishing row between row $n$ and row $m$.
Therefore the  determinant  vanishes as well. 
The time derivative of the determinant is again given by Eq. (\ref{detderiv})
Since we know that in the matrix $\Omega$ there are rows with all zeros
the only non zero contribution to the determinant of the derivative of $\Omega$
can originate out of cofactors with index corresponding to a vanishing row.
But these cofactors in turn will contain either a vanishing row or 
two rows differing just by a multiplicative factor.
Therefore also the derivative of the determinant vanishes.

Thus, we have completed the demonstration that a vanishing quantum Entanglement
gives both a vanishing determinant and a vanishing time derivative
of the determinant, in case of  pure  initial state. This means
the two following relations are true:
\BEA
\label{zeroEntgivesMdM0}
if \; \; E\left(\hro \right)=0 \;\Rightarrow
\;\Delta(t^*)=
\frac{d}{dt}\Delta(t^*)=0&&\\
if \; \Delta(t^*) \neq 0
\; or \;
\frac{d}{dt}\Delta(t^*)\neq 0 \;\Rightarrow
E\left(\hro \right)&\neq0&
\label{MdMnnvanGivesEnnvanish}
\EEA


{\it{The case of  initial mixed state.}}  Let us go back to the two q-bits
case and consider the case where the time evolution is driven by the following Hamiltonian:
\begin{equation}
\hat{H}=\sum_{i=1}^4 E_i |E_i \rangle \langle E_i|,
\label{partham}
\end{equation}
whose eigenvalues and eigenkets are (in units with $\hbar=1$):
\begin{equation}
 E_1=4,\hspace{1em} E_2=2,\hspace{1em}E_3=1,\hspace{1em}
 E_4=0,
 \label{parteigvals}
 \end{equation}
 \BEA
|E_1\rangle&=& |+\rangle_z|-\rangle_z \; ,\hspace{0.5em}
|E_2\rangle=\frac{1}{\sqrt{2}}\left( |+\rangle_z|+\rangle_z
|-\rangle_z|+\rangle_z\right)\\
\nonumber |E_3\rangle&=&
|-\rangle_z|-\rangle_z\; ,\hspace{0.5em} |E_4\rangle=\frac{1}{\sqrt{2}}\left(
|-\rangle_z|+\rangle_z -|+\rangle_z|+\rangle_z\right),
\label{parteigvects}
\EEA
where $ |\pm\rangle_z$ are the eigenkets of the $z$-component of
the Spin operator. Let us consider, now, the 
case in which the measurement is performed on the $1/2$-Spin operators
$\hat{O}_S=O^{0}_{S} \hone+\sum_{i=1}^3 O^i_S \si_i$ and $\hO_A=O^0_{A}
\hone+\sum_{j=1}^3 O^j_S \hat{s}_j$. We observe the time
evolution driven by the Hamiltonian (\ref{partham}) in two
particular cases in which the observables and the initial state of the system
are described by the two following set of  parameters:
\BEA
\label{i1}
\nonumber &&|O^{1}_{A}|>|O^2_A|,\hspace{0.1em}
O^1_S=O^2_S=O^3_S=1\\
&&  \lambda_1=0,\hspace{0.1em}\lambda_2=\lambda_3=1/4,\\
&& \nonumber \rho_1=1/\sqrt{2},\hspace{0.1em}\rho_2=1/\sqrt{2},\hspace{0.1em}
\rho_3=0,
\EEA
corresponding to the spin of interest initially described by a pure state,
and
\BEA
\label{i2}
\nonumber &&|O^1_A|>|O^2_A|,\hspace{0.1em}O^1_S=O^2_S=O^3_S=1,\\
 && \lambda_1=0,\hspace{0.1em}\lambda_2=\lambda_3=1/4,\\
 && \nonumber \rho_1=1/3,
\hspace{0.1em}\rho_2=1/4,\hspace{0.1em} \rho_3=1/2,
\EEA
corresponding to the spin of interest  initially described by a mixed state,
with the ancilla in a mixed state in both cases.
After some long but straightforward algebra, we find out that in both cases (\ref{i1}) and (\ref{i2}), at the time instant
$t^{\ast}=\pi/2$, the Entanglement of Formation, \cite{woters},
\emph{vanishes}, while the Determinant \emph{does not}, $\Delta(t^*)=3 \left( O^A_2 O^A_2 -O^A_1 O^A_1 \right) /
128>0 $. Thus, in general, when the total sistem is initially described by a
mixed quantum state,  properties 
 (\ref{zeroEntgivesMdM0}) and (\ref{MdMnnvanGivesEnnvanish})  are  not
true.
\begin{figure}[bhb]
\vspace{0.35 cm}
\includegraphics[height=4.25 cm, width=7.25 cm]{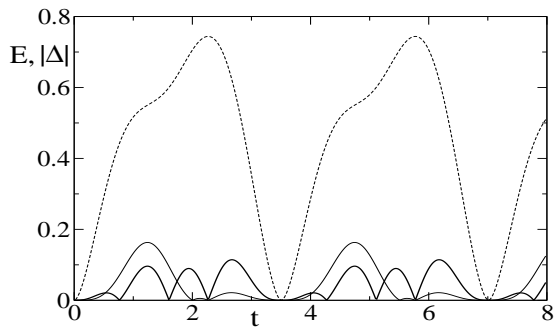}
\vspace{0.2cm}
\caption{Entanglement $E$ (dashed line) and absolute value of the determinant
$|\Delta|$   for two different choices
of commuting observables (normal and thick continuous line) vs. time
(dimensionless units).  Both the
system $S$ and the ancilla $A$  are in an initial pure state as given by
Eq. (\ref{initial})  with $\rho_1=\rho_2=0,\rho_3=1, \lambda=1$. The
interaction is given by  Eq. (\ref{interacteo})  with $\cos(2 \phi)=1/\sqrt{3}$
as in \cite{teo}.}
\label{fig_00}
\end{figure}
\begin{figure}[bhb]
\vspace{0.4 cm}
\includegraphics[height=4.25 cm, width=7.25 cm]{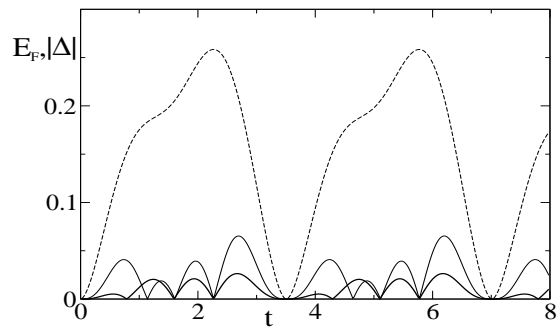}
\vspace{0.2cm}
\caption{ Entanglement of formation $E_F$ (dashed line) and  absolute value of
the determinant $|\Delta|$ for two different
choices of commuting observables (normal and thick continuous line) vs. time
 (dimensionless units).
 System $S$ is initially  in a pure state and the ancilla $A$  in a  mixed
state, as given by Eq. (\ref{initial}) with $\rho_1=\rho_2=0,\rho_3=1,
\lambda=0.5$. The interaction is given by
Eq. (\ref{interacteo}) with $\cos(2 \phi)=1/\sqrt{3}$  as in
\cite{teo}. }
\label{fig_01}
\end{figure}

It is obvious that a null entanglement at any time, which implies a factorized
condition, implies as well a zero determinant for the matrix $M$
and therefore a condition in which the protocol to measure an initial
unknown quantum state here discussed, is not feasible.
The reconstructability of the $1/2$ spin of interest depends on the
quantum Entanglement with the Ancilla system, generated by the time
evolution, so, we would expect   a \emph {non-vanishing} quantum
Entanglement to be \emph{related} to a \emph{non-vanishing}
determinant $\Delta(t)$. Surprisingly, the
relation is not so straightforward: \emph{only} in the  case of an
\emph{initially pure } state, a \emph{vanishing quantum
Entanglement} gives both a \emph{vanishing determinant} and  a
\emph{vanishing time derivative of the determinant}.
This means that in those instants of time when  entanglement vanishes not only the initial state
cannot be reconstructed  but also  that the reconstruction process remains
inefficient in immediate future and past times, since the determinant is zero
to first order included. So the  intervals of time around a time of vanishing
entanglement  must be avoided in order to have an efficient reconstruction process of the
initial state.

In Figs 1 and 2 we  plot  the entanglement and the determinant for the case
of initial pure state and mixed state respectively, adopting for a pure 
state the standard definition of entanglement $E[\rho_T(t)]$
\BEQ
\nonumber
E[\rho_T(t)]=-Tr[\rho_S(t)\log_2 \rho_S(t)]=-Tr[\rho_A(t)\log_2\rho_A(t)]
\EEQ
with $\rho_S(t)$ and $\rho_A(t)$ the partial traces over the system $S$
and the ancilla $A$ respectively.
For the mixed state we adopt as entanglement measure the
so called "Entanglement of Formation" $E_F$ as originally introduced in \cite{woters}.
The results in both figures refer to the case of  $S$ and $A$ interacting
through the following operator:
\BEA \label{interacteo}
\hat{H}&=&  \frac{\hat{\sigma_1}}{\sqrt{2}} \otimes(\cos(\phi)
\hat{s}_2+\sin(\phi) \hat{s}_3)\\
\nonumber &+&\hat{1}\otimes \frac{1}{2}[(\hat{s}_2-\hat{s_1})\sin(\phi)+\hat{s_3} \cos(\phi)]
\EEA
as assumed in \cite{teo}.

{\it{Inverse implication.}}  We wish, now, to study the validity of the
inverse implication of (\ref{zeroEntgivesMdM0}).
To this purpose, we assume that, at a certain instant $t^{\ast}$,
the system is described by a quantum state whose Entanglement does
not vanish. If we find out that, at least, either the determinant or
its time derivative does not vanish, the inverse of implications
(\ref{zeroEntgivesMdM0})  and (\ref{MdMnnvanGivesEnnvanish}) is
proved. Thus, let us assume that, at a certain instant $t^{\ast}$,
the whole quantum system is described by a \emph{pure} state,
$|\Psi  \rangle$, whose quantum Entanglement
does not vanish. We remind that an orthonormal base set
$\left\{|e_i\rangle, i=1,2\right\}$ of the Hilbert space
$\mathcal{H}_S$ and a orthonormal base set $\left\{|f_i\rangle,
i=1,2\right\}$ of the Hilbert space $\mathcal{H}_A$ do exist, such
that the following relation holds true: 
\BEQ
|\Psi  \rangle=\sum_{i=1}^2\sqrt{\lambda_i} |e_i\rangle
\otimes |f_i\rangle
\EEQ
 which is the Schmidt \cite{SchmidtPF} polar form of the
quantum state ket $|\Psi \rangle$. Obviously,
since the quantum Entanglement of
$|\Psi \rangle$ does not vanish, the reduced
density matrix has no vanishing eigenvalues, which means:
$0<\lambda_1=1-\lambda_2<1$.

We consider the Schmidt polar form of the state ket $|\Psi
 \rangle$ and we evaluate the expectation values
of the operators $\hat{O}_S  ,
\hat{O}_A $ and $\left(\hat{O}_S \otimes \hat{O}_A\right) $,
 given by the following expressions:
\BEA\label{avO1O2Ennvanish}
\nonumber \langle \hat{O}_S  \rangle_{t^*}&=&\sum_{j=1}^2 \lambda_j
\langle e_j | \hat{O}_S | e_j\rangle\\
\quad \langle
\hat{O}_A  \rangle_{t^*}&=&\sum_{j=1}^2 \lambda_j \langle f_j
| \hat{O}_A | f_j\rangle,\\
\nonumber  \langle   \hat{O}_S \otimes \hat{O}_A 
 \rangle_{t^*} &=&\sum_{j,k=1}^{2} \sqrt{\lambda_j
\lambda_k}\langle e_j | \hat{O}_S | e_k\rangle \langle f_j
| \hat{O}_A | f_k \rangle. 
\EEA
 Let us consider, now, the particular
case where the observables are described by the following hermitian
operators: $\hat{O}_S=\omega_1 \left(|e_1\rangle\langle
e_2|+|e_2\rangle\langle e_1|\right) , \quad \hat{O}_A=\omega_2
\left( |f_1\rangle \langle f_2|+|f_2\rangle\langle f_1|\right)$.
Starting from Eq.
(\ref{avO1O2Ennvanish}), we easily get the following useful
equalities: 
\BEA
 \langle \hat{O}_S  \rangle_{t^*}&=&\langle \hat{O}_A
 \rangle_{t^*}=0\\
\nonumber \langle  \hat{O}_S\otimes\hat{O}_A 
  \rangle_{t^*}&=&2\sqrt{\lambda_1\left(1-\lambda_1\right)}\omega_1 \omega_2,
\EEA
 which means that both the first and the
second row of the matrix $\Omega$ vanish; thus,
according to the relation (\ref{detderiv})
 both the determinant and its time derivative vanish at the
instant $t^{\ast}$. So, we have demonstrated that properties
described by the \emph{inverse} of implications
(\ref{zeroEntgivesMdM0}) and (\ref{MdMnnvanGivesEnnvanish}), are
\emph{not true}. We stress that the  Schmidt polar form depends on
the instant $t^{\ast}$, so we need to know the time evolution of
the initial state ket in order to find out the particular operators
$\hat{O}_S$ and $\hat{O}_A$ involved in the above demonstration. We also
stress a cue point:  in case of \emph{mixed} states, every measure of the quantum Entanglement has to
vanish for separable mixed states, i.e. for any ensemble of
 bipartite factorized quantum states; thus, our results hold
true for \emph{every } measure of the quantum Entanglement.

 In conclusion we have considered a protocol 
for the determination of an unknown quantum state of a system $S$ 
based on the interaction with an ancilla system $A$, as originally proposed in \cite{teo}.
This protocol allows to determine the initial quantum state of systems $S$
with a single measurement apparatus.
Starting from a factorized condition, it is obvious that an interaction  entangling
the systems $S$ and $A$ is necessary  for the protocol to work.
Therefore it is natural to think that a connection between Entanglement
and the determinant of the linear transformation connecting the parameters
individuating the initial quantum state of the system $S$ to the measurement
of three (commuting) observables at a later time $t^*$ should exist.
We find that in the case of initial pure state of both $S$ and $A$ 
a vanishing entanglement individuates those intervals of time at which 
the reconstruction process is least efficient.
This relation is lost in the case the ancilla system is prepared in an initial
mixed state.
It is rather surprising that, in the case of a mixed state,   even if at a given time $t^*>0$ 
the total density matrix is again separable, interactions exist such that the
initial quantum state of the system $S$ can still be recovered from
measurements done at this time, and  we have provided an example of such interactions.
This seems to aim at the long debated  different nature  between mixed and pure
quantum states and  at the different physical meaning of entanglement in the
two cases.

\begin{acknowledgments}
We thank dr. Pasquale Calabrese for critical reading of this manuscript.
\end{acknowledgments}


\begin{thebibliography}{99}




\bibitem{chuang}I.L. Chuang {\it et al.} Nature (London) {\bf 393}, 143 (1998)

\bibitem{hillery}
J. A. Bergou, U Herzog and M. Hillery, Phys. Rev. A. {\bf 71},
042314 (2005).

\bibitem{bennett}
C. H. Bennett and D. P. Divincenzo, Nature (London) {\bf 404}, 247 (200)

\bibitem{teo} Armen E. Allahverdyan, R. Balian and  Th. M. Nieuwenhuizen,
Phys. Rev. Lett. {\bf 92} 120402 (2004)

\bibitem{ariano}G. M. D'Ariano, Phys. Lett. A {\bf 300},1 (2002).
\bibitem{du}
 Jiangfeng Du, Min Sun, Xinhua Peng and Thomas Durt
Phys. Rev. A {\bf 74} 042341 (2006)

\bibitem{woters}
 W. K. Wootters Phys.Rev.Lett. {\bf 80} 2245 (1998) 

\bibitem{SchmidtPF} E. Schmidt, Math Ann. 63 (1906) 433

\bibitem{mine1}G. Aquino and  B. Mehmani,
Proceedings of the Workshop "Beyond the Quantum", pp. 115-124, (World
Scientific 2007).

\bibitem{bahar2}
B. Mehmani, A. Allahverdyan and Th. M. Neuwenhuizen,
Phys. Rev. A  {\bf 77}, 032122 (2008)

\end{thebibliography}
\end{document}